\documentclass{vldb}
\usepackage{graphicx}
\usepackage{balance}  

\usepackage{booktabs} 

\usepackage{algorithm}
\usepackage{algorithmic}

\usepackage{url}

\usepackage{array}
\newcommand{\PreserveBackslash}[1]{\let\temp=\\#1\let\\=\temp}
\newcolumntype{C}[1]{>{\PreserveBackslash\centering}p{#1}}
\newcolumntype{R}[1]{>{\PreserveBackslash\raggedleft}p{#1}}
\newcolumntype{L}[1]{>{\PreserveBackslash\raggedright}p{#1}}

\begin{document}


\title{Understanding tree: a tool for estimating one's understanding of conceptual knowledge}



%
%
%
%

\numberofauthors{1} 

\author{
%
%
\alignauthor
Gangli Liu\\
        \affaddr{Tsinghua University}\\
        \affaddr{Beijing 100084, China}\\
        \email{gl-liu13@mails.tsinghua.edu.cn}
}


\maketitle

\begin{abstract}
People learn whenever and wherever possible, and whatever they like or encounter--Mathematics, Drama, Art, Languages, Physics, Philosophy, and so on. With the bursting of knowledge, evaluation of one's understanding of conceptual knowledge becomes increasingly difficult. There are a lot of demands for evaluating one's understanding of a piece of knowledge, e.g., facilitating personalized recommendations; discovering one's expertises and deficiencies in a field; recommending topics for a conversation between people with different educational or cultural backgrounds in their first encounter; recommending a learning material to practice a meaningful learning etc. Assessment of understanding of knowledge is conventionally practiced through tests or interviews, but they have some limitations such as low-efficiency and in-comprehensive. We propose a method to estimate one's understanding of conceptual knowledge, by keeping track of his/her learning activities. It overcomes some limitations of traditional methods, hence complements traditional methods. 
\end{abstract}

\section{Introduction}

Our world is bursting with knowledge. Nearly every discipline has been subdivided into numerous sub-disciplines.  People learn whenever and wherever possible. People's learning of knowledge is not confined to childhood or the classroom but takes place throughout life and in a range of situations; it can take the form of formal learning or informal learning \cite{Paradise2009}, such as daily interactions with others and with the world around us. Lifelong learning is the ``ongoing, voluntary, and self-motivated" pursuit of knowledge for either personal or professional reasons \cite{Cliath2000}. According to Tough's study, almost 70\% of learning projects are self-planned \cite{Tough1979}. 

As people learn eternally, one significant issue is to evaluate how much knowledge an individual is possessing at a particular time. E.g., suppose we have a database that records all the entries of Wikipedia and a person's understanding degree of each entry (e.g., on a scale of 1 to 10). With this information, a lot of new applications are becoming practical. The following are some examples:

\subsubsection*{1. Determine a person's knowledge state}
If we set a threshold to the understanding grades, and assume the subject has understood a knowledge entry if its grade is larger than the threshold, then we can determine a person's knowledge state and knowledge composition at a particular time. Like Figure 1.

\begin{figure}
	\centering
	\includegraphics[width=0.8\columnwidth]{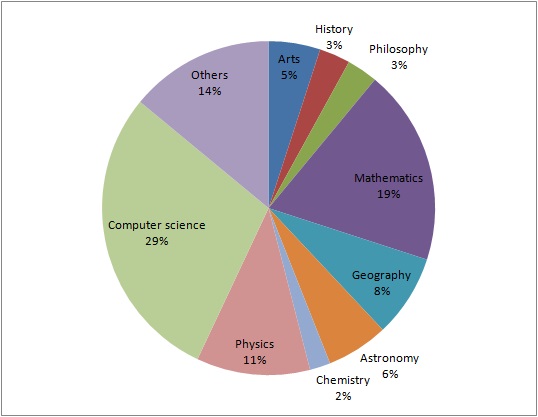}
	\caption{A person's knowledge composition (imaginary)}
\end{figure}

\subsubsection*{2. Discover a person's expertises and deficiencies}	
Expertise finding is critical for an organization or project. Since the participation of experts plays an important role for the success of an organization or project. With the understanding evaluation database, it is convenient to discover a person's domain-level expertise and topic-level expertise. E.g., if a person has a question of what is Poincare Conjecture, he do not need to ask a mathematician (domain-level expert), who is not always available. Instead, he can ask one of his friends who is not a mathematician but has topic-level expertise in it. Besides expertise, we can also discover a person's deficiencies in a field, so he can remedy the deficiencies.

\subsubsection*{3. Make personalized recommendations}
If we know a person's understanding degree to each piece of knowledge, it is natural to make personalized recommendations to the subject based on the information. E.g., suppose we know a person has good understanding in the topics of Deep Learning, then we can recommend latest papers about Deep Learning to the subject, or recommend the person as a reviewer of a paper related to Deep Learning. In addition, we can recommend learning materials to the subject to help him practice meaningful learning, which will be discussed in details in Section 4.

\subsection{Procedural and Conceptual Knowledge}
Studies of knowledge indicate that knowledge may be classified into two major categories: procedural and  conceptual knowledge \cite{hiebert2013conceptual,mccormick1997conceptual}. Procedural knowledge is the knowledge exercised in the accomplishment of a task, and thus includes knowledge which cannot be easily articulated by the individual, since it is typically nonconscious (or tacit). It is commonly referred to as ``know-how". Such as knowing how to cook delicious food, how to drive an airplane, or how to play basketball etc. Conceptual knowledge is quite different from procedural knowledge. It involves understanding of the principles that govern a domain and of the interrelations between pieces of knowledge in a domain; it concerns understanding and interpreting concepts and the relations between concepts. It is commonly referred to as ``know-why", such as knowing why something happens in a particular way. In this article, we only deal with evaluating a person's understanding of conceptual knowledge.

\subsection{The framework}
At present, assessment of one's understanding of conceptual knowledge is primarily through tests \cite{pisa2000measuring,hunt2003concept,qian2004evaluation} or interviews, which have some limitations such as low-efficiency and in-comprehensive. E.g., it needs other people's cooperation to accomplish the assessment, which is time-consuming; moreover, only a small portion of topics of a domain is evaluated during an assessment, which cannot comprehensively reflect a person's knowledge state of the domain.

We propose a new model called Individual Conceptual Knowledge Evaluation Model (ICKEM) to evaluate one's understanding of a piece of conceptual knowledge quantitatively. It has the advantage of evaluating a person's understanding of conceptual knowledge independently, comprehensively, and automatically. It keeps track of one's daily learning activities (reading, listening, discussing, writing etc.), dividing them into a sequence of learning sessions, then analyzes the text content of each learning session to discover the involved knowledge topics and their shares in the session. Then a learning experience is inserted to the involved knowledge topics' learning histories (It maintains a leaning history for \emph{each} knowledge topic). Therefore, after a period of time (e.g., several years or decades of years), a knowledge topic's leaning history that records the subject's each leaning experience about the topic is generated. Based on the learning history, the subject's familiarity degree to a knowledge topic is evaluated. Finally, it estimates one's understanding degree to a topic, by comprehensively evaluating one's familiarity degrees to the topic itself and other topics that are essential to understand the topic. Figure 2 is the framework of ICKEM. Each hexagon of the diagram indicates a processing step; the following rectangle indicates the results of the process.

\begin{figure}
	\centering
	\includegraphics[width=0.9\columnwidth]{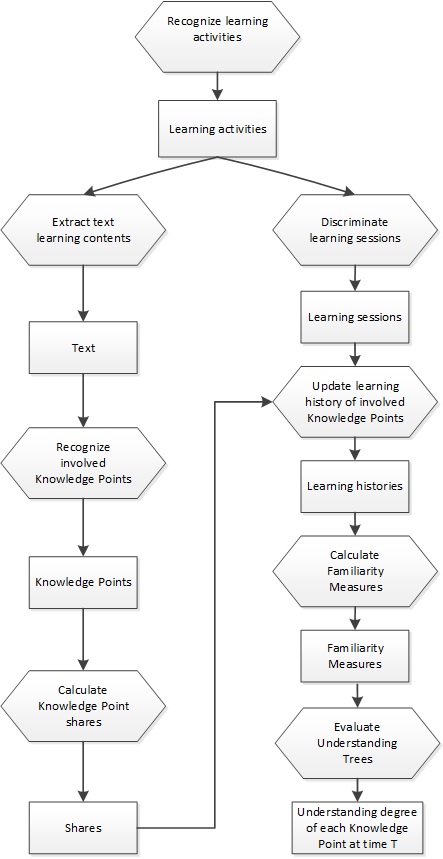}
	\caption{The framework of ICKEM}
\end{figure}

The remainder of this paper is organized as follows. Section 2 discusses how to calculate a person's familiarity degree to a knowledge topic. Section 3 introduces a method to estimate a person's understanding degree to a topic, by checking the familiarity degrees of the topic's understanding tree. It also presents an algorithm to generate a knowledge topic's understanding tree. In Section 4, we introduce an utilization of ICKEM, computer-aided incremental meaningful learning (CAIML), which helps an individual practice meaningful learning. Section 5 discusses related issues about evaluating one's understanding of conceptual knowledge. We cover related work in Section 6, before concluding in Section 7.

\section{Evaluate familiarity degree}
This section introduces the procedures that are devised to evaluate a person's familiarity degree to a knowledge topic. It starts by presenting a formal definition of knowledge and learning, then discusses how to divide a person's daily learning activities into a series of learning sessions, and analyze the text learning content to obtain a topic's share in a session. After these procedures, a knowledge topic's learning history can be generated. Finally, based on the learning history, the subject's familiarity degree to a topic is calculated.

\subsection{Definition of knowledge and learning}
Knowledge is conventionally defined as beliefs that are true and justified. To be `true' means that it is in accord with the way in which objects, people, processes and events exist and behave in the real world. However, exactly what evidence is necessary and sufficient to allow a true belief to be `justified' has been a topic of discussion (largely among philosophers) for more than 2,000 years \cite{hunt2003concept}. In ICKEM, a Knowledge Point is defined as a piece of conceptual knowledge that is explicitly defined and has been widely accepted, such as Bayes' theorem, Euler's formula, mass-energy equivalence, Maxwell's equations, gravitational wave, and the expectation-maximization algorithm etc.

Learning is the process of acquiring, modifying, or reinforcing knowledge, behaviors, skills, values, or preferences in memory \cite{terry2015learning, hunt2003concept, bransford1999people}. An individual's possessing of knowledge is the product of all the experiences from the beginning of his/her life to the moment at hand \cite{hunt2003concept,benassi2014applying}. Learning produces changes in the organism and the changes produced are relatively permanent \cite{schacter2011psychology}.

\subsection{Discriminate learning sessions}
In ICKEM, a person's daily activities are classified into two categories: learning activities and non-learning activities. An activity is recognized as a learning activity if its content involves at least one Knowledge Point of a predefined set. In addition, the learning activities are divided into a sequence of learning sessions, since it is essential to know how many times and how long for each time the individual has learned a Knowledge Point. An individual can employ many methods to learn conceptual knowledge, such as reading, listening, discussing, and writing. Different strategies should be used to discriminate learning sessions for different learning methods. E.g., for leaning by reading documents on a computer, Algorithm 1 is devised to discriminate learning sessions. For other learning methods, it is more complicated to divide learning sessions. However, there are already some attempts for detecting human daily activities \cite{Ni2013,Piyathilaka2013,Sung2012}. Sung et al. devised an algorithm for recognizing human daily activities from RGB-D Images \cite{Sung2012}. They tested their algorithm on detecting and recognizing twelve different activities (brushing teeth, cooking, working on computer, talking on phone, drinking water, talking on a chair etc.) performed by four people and achieved good performance.

Algorithm 1 works by periodically checking the occurrence of the following events: 
\begin{enumerate}
	\item A document is opened;
	\item The foreground window has switched to an opened document from another application (APP), such as a game APP;
	\item After the computer being idled for a period of time, there are some mouse or keyboard inputs detected, which indicates the individual has come back from other things. Meanwhile, the foreground window is an opened document;
	\item A document is closed;
	\item The foreground window has switched to another APP from a document;
	\item The foreground window is a document, but the computer has idled for a certain period of time without any mouse or keyboard inputs detected, the individual is assumed to have left to do other things.
\end{enumerate}

Occurrence of Event 1, 2, or 3 indicates a learning session has started; if Event 4, 5, or 6 occurred, a learning session is assumed being terminated. The duration of a learning session equals the interval between its start and stop time. 

\begin{algorithm}
	\caption{An algorithm to discriminate one's learning sessions when reading.}
	\label{alg1}
	\begin{algorithmic}[1]			
		\WHILE{The Reader APP is running}		
		\IF{Event 4 \textbf{OR} Event 5 \textbf{OR} Event 6 occurred}					
		\STATE Record that a document's learning session has stopped;
		\ELSE
		\IF{Event 1 \textbf{OR} Event 2 \textbf{OR} Event 3 occurred}			
		\STATE Record that a learning session about the current document has started;
		\ENDIF
		\ELSE
		\IF{There is no APP and document switch}			
		\STATE Check and record if there is a Page switch;
		\ENDIF			
		\ENDIF
		\STATE Keep silent for $ T $ seconds;
		\ENDWHILE		
	\end{algorithmic}	
\end{algorithm}

Figure 3 shows some examples of discriminated learning sessions. Attribute ``\textit{did}'' means document ID, which indexes a document uniquely. Attribute ``\textit{actiontype}'' indicates the type of an action. ``\textit{Doc Act}'' means a document has been activated. ``\textit{Page Act}'' is defined  similarly. ``\textit{Doc DeAct}'' means a document has been deactivated. That is to say, a learning session has stopped. Attribute ``\textit{page}'' indicates a page number. Attribute ``\textit{duration}'' records how long a page has been activated in seconds. If two learning sessions' interval is less than a certain threshold, such as 30 minutes, and their learning material is the same (e.g., the same document), they are merged into one session. Therefore, ``Session 2'' and ``Session 3'' are merged into one session.

\begin{figure}
	\centering
	\includegraphics[width=0.8\columnwidth]{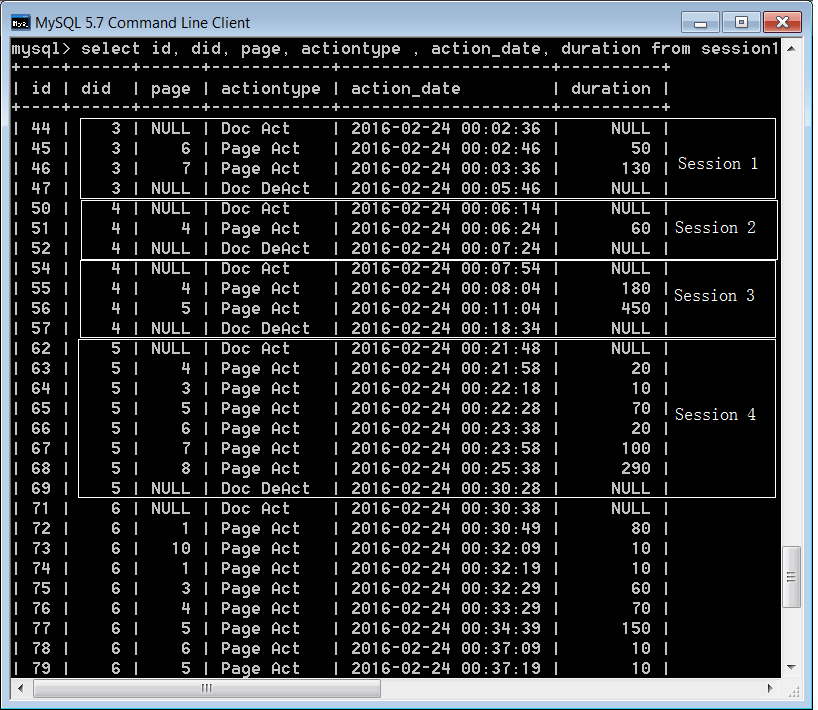}
	\caption{A person's reading learning sessions logged in a database.}
\end{figure}

\subsection{Capture the text learning content}

Most learning processes are associated with a piece of learning material. E.g., reading a book, the book is the learning material; attending a course or discussion, the course and discussion contents can be regarded as the learning material. Some learning materials are text or can be converted to text. E.g., discussing a piece of knowledge with others. The discussion contents can be converted to text by exploiting Speech Recognition. Similarly, if one is reading a printed book, the contents of the book can be captured by wearable computers like Google Glass and then converted to text through Optical Character Recognition (OCR). If the book is electronic, no conversion is needed; the text can be extracted directly. Since Algorithm 1 records the accurate set of pages a person has read during a learning session, only the text content of related pages is extracted.

\subsection{Calculate a Knowledge Point's share}
A learning session may involve many topics, it is necessary to know how much a learning session involves a topic. To calculate a Knowledge Point's share, maybe the simplest way is to deem the text learning content as a bag of words, and calculate a Knowledge Point's share based on its Term Frequency (TF) or normalized TF, like in Equation 1 and 2. $ N_{i} $ is term $ i $'s normalized TF. It is calculated with Equation 1, where $ T_{i} $ is term $ i $'s TF, $ Max(TF) $ is the maximum TF of the captured text content, $ \alpha $ is a constant regulatory factor.

\begin{equation}
	N_{i} = \alpha + (1 - \alpha) * T_{i} / Max(TF)
\end{equation}
\begin{equation}
	\xi_{i} = \frac{N_{i}}{\sum_{j=1}^mN_{j}}
\end{equation}

Another method of discovering a Knowledge Point's share is to analyze the text learning content with topic model. A topic model is a type of statistical model for discovering the abstract "topics" that occur in a collection of documents. Topic model is a frequently used text-mining tool for discovery of hidden semantic structures in a text body \cite{blei2012probabilistic,steyvers2007}. An early topic model was described in \cite{Papadimitriou1998}. Another one, called probabilistic latent semantic analysis (PLSA), was created by Hofmann in 1999 \cite{hofmann1999probabilistic}. Latent Dirichlet allocation (LDA) is the most common topic model currently in use \cite{blei2003latent}. It is a generalization of PLSA. It introduces sparse Dirichlet prior distributions over document-topic and topic-word distributions. Other topic models are generally extensions on LDA.

The inputs of a probabilistic topic model are a collection of $ N $ documents, a vocabulary set $ V $, and the number of topics $ k $. The outputs of a probabilistic topic model are the following:

\begin{itemize}
	\item k topics, each is word distribution : $ \{\theta_{1},...,\theta_{k}\} $;
	\item Coverage of topics in each document $ d_{i} $: $ \{\pi_{i1},...,\pi_{ik}\} $;\\
	$ \pi_{ij} $ is the probability of document $ d_{i}$ covering topic $ \theta_{j} $.	
\end{itemize}

The subject's $ N $ learning sessions'  text-contents can be deemed as a collection of $ N $ documents, then the document collection is analyzed with a topic model like LDA.
Based on the outputs of topic model, Equation 3 can be used to calculate the share of Knowledge Point  $t_{m}$ in learning session  $d_{i}$. $p(t_{m}|\theta_{j})$ is the probability of Knowledge Point  $t_{m}$ in topic  $\theta_{j}$.

\begin{equation}
p(t_{m}|d_{i}) = {\sum_{j=1}^k\pi_{ij}p(t_{m}|\theta_{j})}
\end{equation}

\subsection{The subject's state during a session}
The subject's physical and psychological status may influence the effect of a learning session. E.g., if the subject is tired, or severely injured, the learning effect is usually worse than being in a normal state. Similarly, if the subject is in a state of severe depression, anxiety, or scatterbrained, the learning effect cannot be good either. One issue is how to collect values for these attributes. Fortunately, there are already systems for monitoring a person's physical and psychological status \cite{Yang2005,Cheng2012,DeRemer2015,Kurniawan2013,Lin2014}. Another issue is studies need to be conducted to decide how these attributes would affect the learning effect. Preferable to have some equations to calculate a ``physical and psychological status factor", describing how the learning effect is affected by the subject's physical and psychological state during a session.

On the other hand, if we just want a rough estimation of the subject's familiarity degree to a Knowledge Point, these physical and psychological status attributes may be ignored. Since we are observing the subject in a long time rang, typically several years or decades of years, we can assume the subject is in ``normal'' state most of the time, and ignore its fluctuation.

\subsection{Effectiveness of a learning method}
Different learning methods may have different levels of effectiveness. E.g., learning by reading a book and learning by discussing with others, do these two methods have equivalent levels of memory retention after the learning experience? (supposing other conditions are equivalent, such as learning content, session duration, physical and psychological status etc.) These is no consensus about the effectiveness of different learning methods. The National Training Laboratories (NTL) Institute proposes a \emph{learning pyramid} model, which maps a range of learning methods onto a triangular image in proportion to their effectiveness in promoting student retention of the material taught \cite{sousa2001brain,magennis2005teaching}. However, the credibility and research base of the model have been questioned by other researchers \cite{lalley2007learning,letrud2012rebuttal}. Further research is necessary to make a systematic comparison between the effectiveness of different learning methods. Preferable to have a ``learning method factor" depicting the effectiveness of a method. Similarly, if we just want a rough estimation of the subject's familiarity degrees, we can omit the difference among different learning methods. As \cite{lalley2007learning} concluded, there is no learning method being consistently superior to the others and all being effective in certain contexts. \cite{lalley2007learning} also stressed the importance of reading: reading is not only an effective learning method, it is also the main foundation for becoming a ``life-long learner".

\subsection{A Knowledge Point's learning history}
With the discriminated learning sessions, a Knowledge Point's share in each session, the subject's physical and psychological status during a session, and the ``learning method factor", after a period of time, the subject's learning histories about each Knowledge Point can be generated. Figure 4 shows an exemplary learning history. It records a person's each learning experience about a Knowledge Point. ``LCT" stands for ``learning cessation time". It is used to calculate the interval between the learning time and the evaluation time, which is then used to estimate how much information has been lost due to memory decay. ``Duration" is the length of a learning session. ``Proportion" is the Knowledge Point's share during a learning session. ``PPS factor" stands for the ``physical and psychological status factor". It is a number between 0 and 1 that is calculated based on the subject's average physical and psychological status during a session. ``LM factor" stands for the ``learning method factor". It is also a number between 0 and 1 that is allocated to a learning method according to its effectiveness level.

\begin{figure}
	\centering
	\includegraphics[width=1.0\columnwidth]{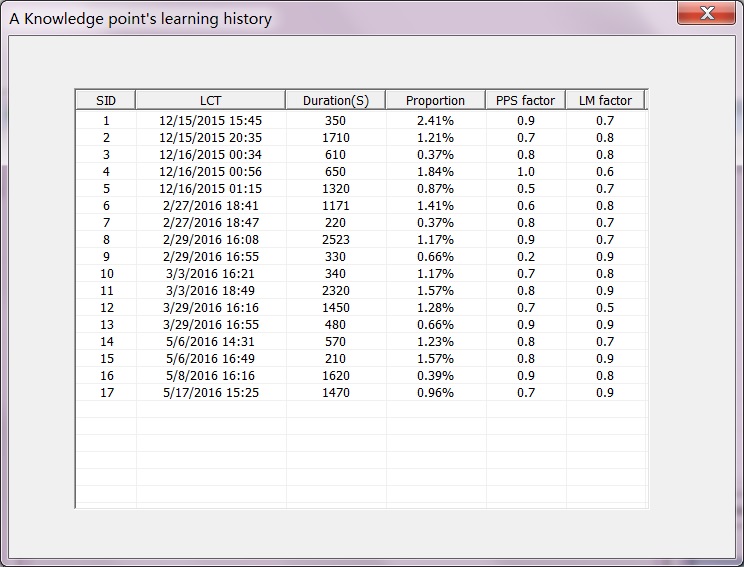}
	\caption{An person's learning history of a Knowledge Point}
\end{figure}

\subsection{Memory retention of a learning experience}
People learn all the time; meanwhile, people forget all the time. Human memory declines over time. Interestingly, most researchers report there is no age difference for memory decay \cite{rubin1996one}. To calculate the familiarity degree, we need to address how the effect of a learning experience decays over time. However, there is no consensus of how human memory decays. Psychologists have suggested many functions to describe the monotonic loss of information with time \cite{rubin1996one}. But there is no unanimous agreement of which one is the best. The search for a general description of forgetting is one of the oldest unresolved problems in experimental psychology \cite{averell2011form}. 

We propose to use Ebbinghaus' forgetting curve equation \cite{ebbinghaus1913memory} to describe the memory retention of a learning experience over time. Since it is the most well-known forgetting curve equation and its soundness has been proved by many studies \cite{murre2015replication,finkenbinder1913curve,heller1991replikation}.
Ebbinghaus found Equation 4 can be used to describe the proportion of memory retention after a period of time\footnote{It can be found at \url{http://psychclassics.yorku.ca/Ebbinghaus/memory7.htm}}, where $ t $ is the time in minutes counting from one minute before the end of the learning, $ k $ and $ c $ are two constants that equal 1.84 and 1.25, respectively. Figure 5 shows the percentage of memory retention over time calculated by Equation 4. The Y axis is the percentage of memory retention; the X axis is the time-since-learning in minutes. It can be seen that memory retention declines drastically during the first 24 hours (1,440 minutes), then the speed tends to be steady.

\begin{equation}
b(t) = k/((\log t)^c + k)
\end{equation}

\begin{figure}
	\centering
	\includegraphics[width=0.8\columnwidth]{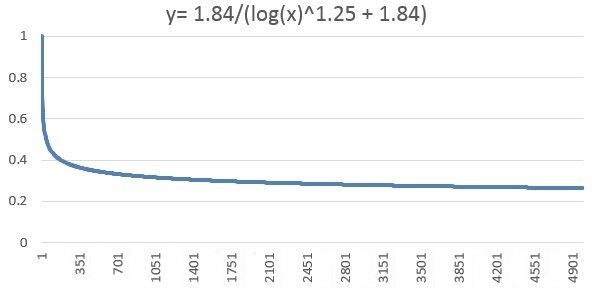}
	\caption{The percentage of memory retention over time calculated by Equation 4}
\end{figure}

\subsection{Calculate the familiarity degree}

Based on the learning history, Equation 5 is utilized to calculate the subject's Familiarity Measure to a Knowledge Point  $ k_{i} $ at time $ t $. A Familiarity Measure is defined as a score that depicts a person's familiarity degree to a Knowledge Point. If the unit of time is the second, its unit is defined as \emph{gl}. The input is $ k_{i} $'s learning history--a sequence of $ m $ learning sessions (like Figure 4). $ d_{j} $ is session $ j $'s duration; $ \xi_{ij} $ is Knowledge Point $ k_{i} $'s share in session $ j $; $ t_{j} $ is session $ j $'s ``learning cessation time", $ b(t-t_{j}) $ calculates the percent of memory retention of session $ j $ at time $ t $ with Equation 4; $ F^{pps}_{j} $ is the ``physical and psychological status factor" of session $ j $; $ F^{lm}_{j} $ is the ``learning method factor" of session $ j $.

\begin{equation}
	f(k_{i},t) = \sum_{j=1}^md_{j}*\xi_{ij}*b(t-t_{j})*F^{pps}_{j}*F^{lm}_{j}
\end{equation}

The computation hypothesizes each learning experience about a Knowledge Point contributes some effect to the subject's current understanding of it, and the learning effect declines over time according to Ebbinghaus' forgetting curve. Other attributes (the subject's physical and psychological status, learning method) that may affect the learning effect are counted in as numeric factors. The Familiarity Measure of a Knowledge Point is calculated as the additive effects of all the learning experiences about it. If we omit the influence of subject physical and psychological status and the difference among learning methods, Equation 5 can be simplified to Equation 6.

\begin{equation}
	f(k_{i},t) = \sum_{j=1}^md_{j}*\xi_{ij}*b(t-t_{j})
\end{equation}

Table 1 compares Familiarity Measures calculated by Equation 5 and Equation 6 at different times based on the learning history of Figure 4. The evaluation times are an hour, a day, a month, a year, and 10 years after last learning of the knowledge point.

\begin{table*}
	\centering
	\begin{tabular}{ c  c  c  c c c }
		\hline
		& 5/17/2016 16:25 & 5/18/2016 15:25 & 6/17/2016 15:25 &5/17/2017 15:25 & 5/17/2026 15:25\\ \hline
		Equation 5 & 25	&23.5&	21.9&	19.4&	16.6\\ \hline		
		Equation 6 & 42.6&	40.2&	37.6	&33.3	&28.5\\  \hline
	\end{tabular}
	\caption{Familiarity Measures calculated at different times}
\end{table*}

\section{Estimate understanding degree}
Understanding is quite subtle to measure. We hypothesize that if a person is familiar with a Knowledge Point itself and all the background knowledge that is essential to understand it, he should have understood the Knowledge Point. Because Familiarity Measure depicts the cumulated effects of one's learning experiences about a topic, high levels of Familiarity Measures imply intensive learning activities about the suite of knowledge topics. Intensive learning activities usually result in a good understanding.

The background Knowledge Points that are essential to understand a Knowledge Point can be extracted by analyzing its definition. Table 2 lists eight reduced documents, each of them is a definition of a Knowledge Point in Probability Theory or Stochastic Process, the texts are quoted from Wikipedia and other websites. The third column of Table 2 lists the involved Knowledge Points in the documents, which are deemed as the background knowledge to understand the host Knowledge Point.

\begin{table*}
	\centering
	\begin{tabular}{|C{1cm}|L{12cm}|C{2.5cm}|}
		\hline
		Doc & \qquad \qquad \qquad \qquad \qquad \qquad \qquad \qquad Content & Knowledge Points  \\ \hline
		D1 & A Strictly Stationary Process (SSP) is a Stochastic Process (SP) whose Joint Probability Distribution (JPD) does not change when shifted in time. &  SSP, JPD,\qquad\qquad Time, SP \\ \hline
		\qquad \qquad \qquad \qquad\qquad \qquad D2 &  A Stochastic Process (SP) is a Probability Model (PM) used to describe phenomena that evolve over time or space.  In probability theory, a stochastic process is a Time Sequence (TS) representing the evolution of some system represented by a variable whose change is subject to a Random Variation (RaV).  &  SP, PM, TS, Time, Space, System,\qquad \qquad Variable, RaV \\ \hline
		\qquad \qquad \qquad\qquad \qquad \qquad D3 &  In the study of probability, given at least two Random Variables (RV) X, Y, ... that are defined on a Probability Space (PS), the Joint Probability Distribution (JPD) for X, Y, ... is a Probability Distribution (PD) that gives the probability that each of X, Y, ... falls in any particular range or discrete set of values specified for that variable.  &  JPD, RV,\qquad\qquad PS, PD,\qquad \qquad Variable, Probability \\ \hline
		\qquad \qquad D4 &  A Probability model (PM) is a mathematical representation of a random phenomenon. It is defined by its Sample Space (SS), events within the SS, and probabilities associated with each event.  &  PM, SS,\qquad\qquad Event, Probability \\ \hline
		
		D5 &  In probability and statistics, a Random variable (RV) is a variable quantity whose possible values depend, in some clearly-defined way, on a set of random events.  &  RV, Variable, Event \\ \hline
		\qquad \qquad D6 &  A Probability Space (PS) is a Mathematical Construct (MC) that models a real-world process consisting of states that occur randomly. It consists of three parts: a Sample Space (SS), a set of events, and the assignment of probabilities to the events.  &  PS, MC, SS, Probability, Event \\ \hline
		D7 & A Probability Distribution (PD) is a table or an equation that links each outcome of a statistical experiment with its probability of occurrence. &  PD,\qquad\qquad Probability \\ \hline
		D8 & The Sample Space (SS) is the set of all possible outcomes of the samples. &  SS, Sample \\  	 
		\hline 
	\end{tabular}
	\caption{A list of documents and their involved Knowledge Points}
\end{table*}

An Understanding Tree is a treelike data structure which compiles the background Knowledge Points that are essential to understand the root Knowledge Point. Figure 6 illustrates four Understanding Trees based on the definitions of Table 2. The nodes of the tree can be further interpreted by other Knowledge Points until they are Basic Knowledge Points (BKP). A BKP is a Knowledge Point that is simple enough so that it is not interpreted by other Knowledge Points. Figure 7 shows a fully extended Understanding Tree based on the definitions of Table 2. Figure 8 shows an exemplary Understanding Tree with all the redundant nodes eliminated. Each node is tagged with an artificial Familiarity Measure, which can be calculated with Equation 5 or 6 in practice. The leaf nodes of Figure 7 and Figure 8 are BKPs. The height and number of nodes of an Understanding Tree characterize the complexity degree of it. Understanding Tree can be used to evaluate a person's topic-level expertise, such as evaluating a person's understanding to a specific algorithm like the Quicksort algorithm.

\begin{figure}
	\centering
	\includegraphics[width=0.99\columnwidth]{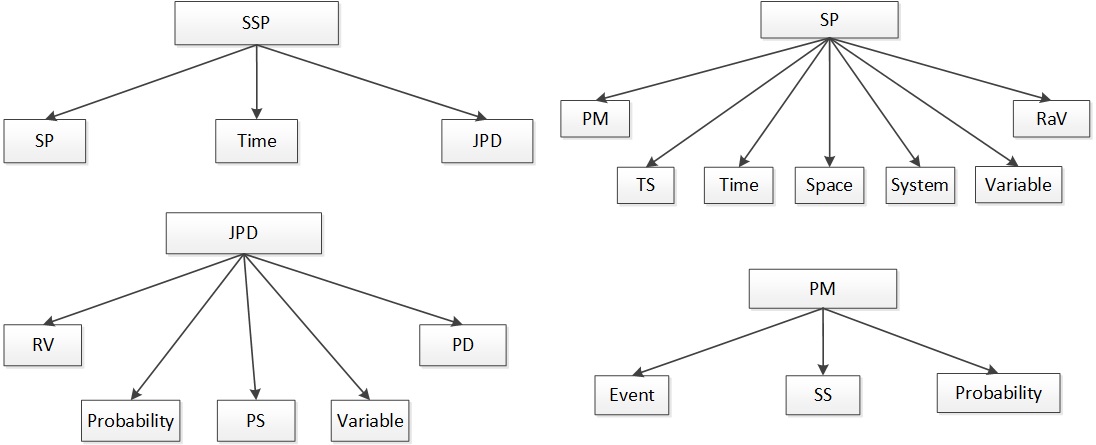}
	\caption{Some examples of Understanding Trees}
\end{figure}

\begin{figure*}
	\centering
	\includegraphics[width=1.3\columnwidth]{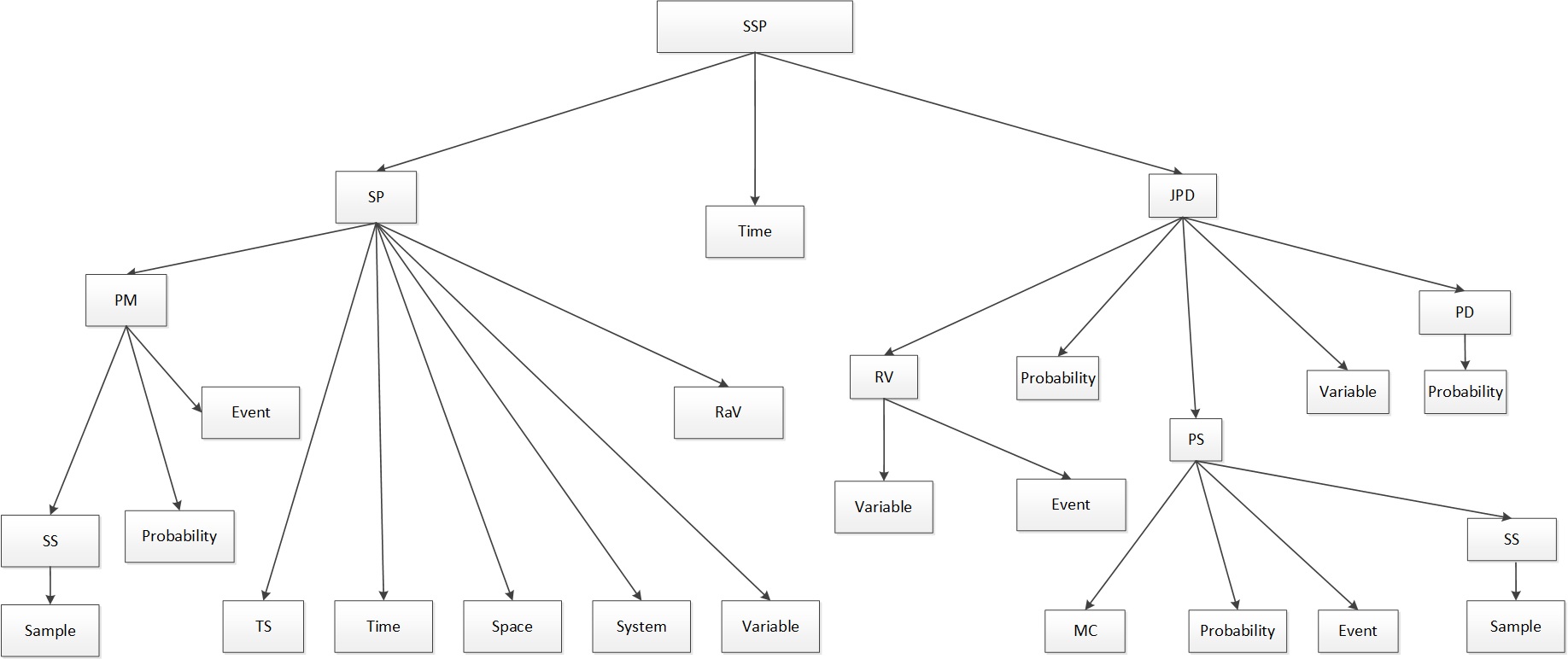}
	\caption{A fully extended Understanding Tree}
\end{figure*}

\begin{figure*}
	\centering
	\includegraphics[width=1.3\columnwidth]{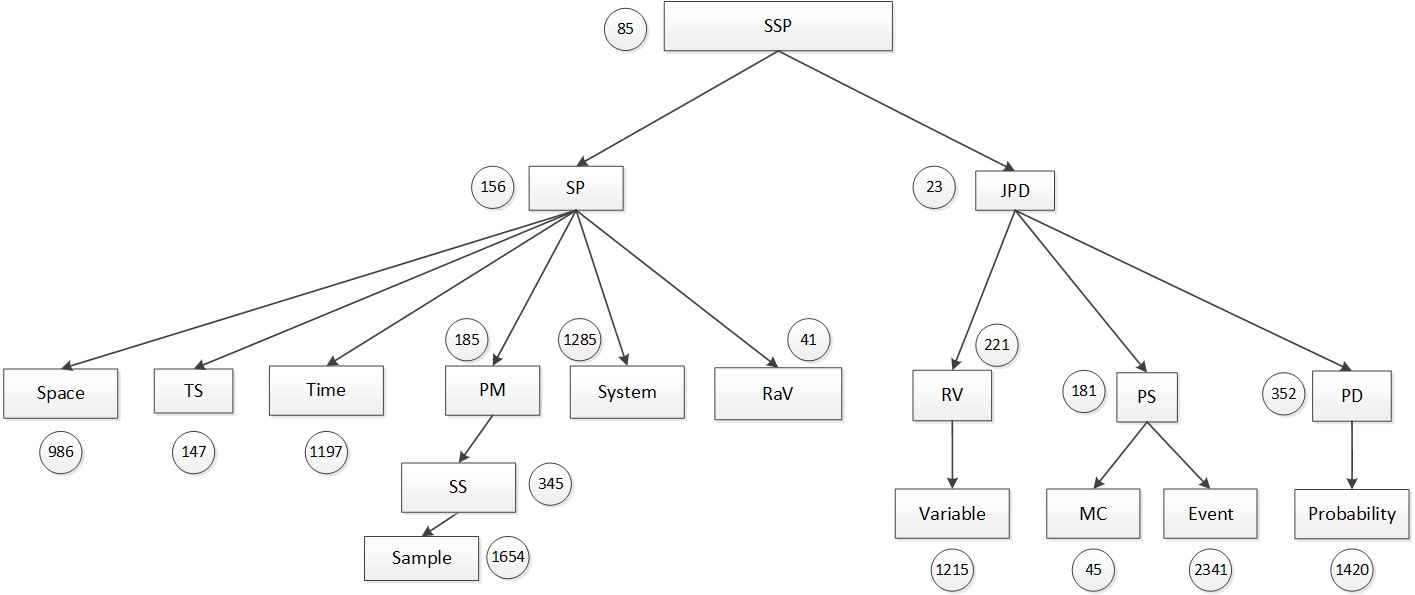}
	\caption{A standard Understanding Tree tagged with Familiarity Measures}
\end{figure*}

\begin{figure*}
	\centering
	\includegraphics[width=1.3\columnwidth]{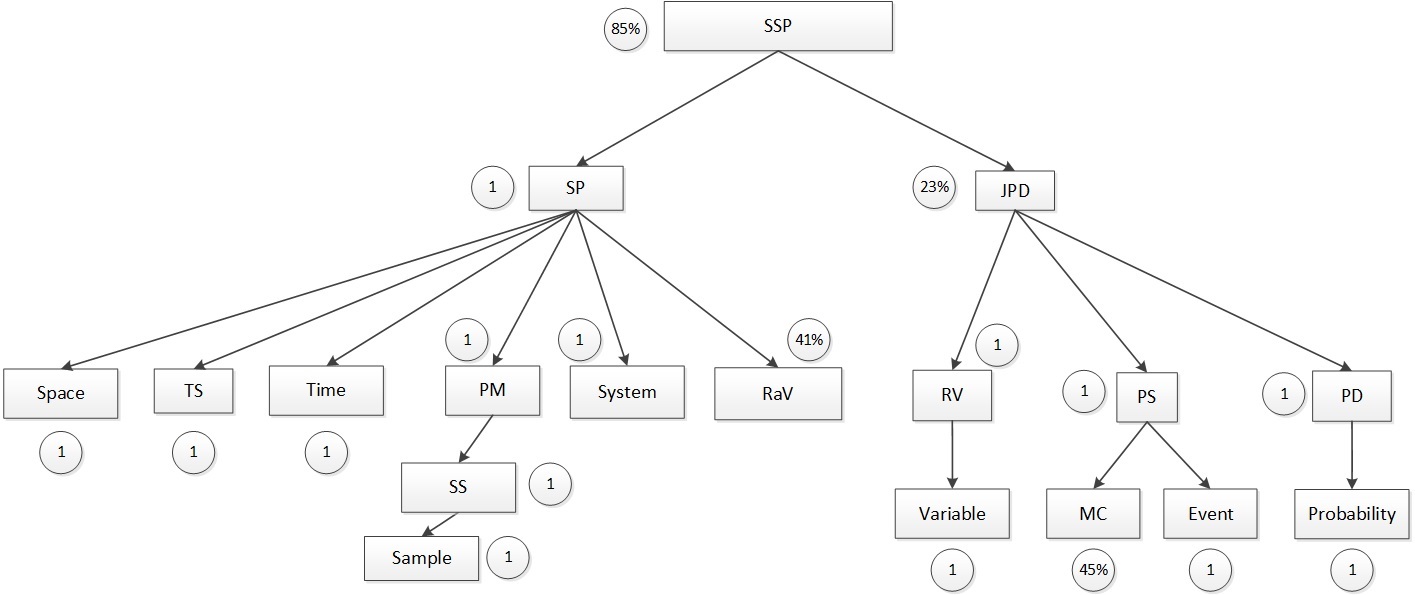}
	\caption{Familiarity Measures transformed into percentages}
\end{figure*}

\subsection{Calculation of understanding degree}
If all the Familiarity Measures of an Understanding Tree exceed a threshold (such as 100), it is assumed that the person has understood the root Knowledge Point. Then the Knowledge Point is classified as ``Understood''; otherwise, it is classified as ``Not Understood''. Due to the differences of people's intelligence and talent, different people may have different thresholds. 

If a Familiarity Measure is less than the threshold, a percentage is calculated by dividing it by the threshold, indicating the subject's percent of familiarity of the node; if the Familiarity Measure is greater than the threshold, the percentage is set to 1, implying the quantity of familiarity of this node is large enough for understanding the root, extra familiarity is good but not necessary. Equation 7 is used to calculate the percentage of familiarity, $ f(k_{i},t) $ is the subject's Familiarity Measure to Knowledge Point $ k_{i} $ at time $ t $, $ f_{T} $ is the threshold.

If a Knowledge Point is classified as ``Not Understood'', a percent of understanding is calculated with Equation 8. $ P_u(k_{r},t) $ is the subject's percent of understanding of the root Knowledge Point at time $ t $, $ P_f(k_{r},t) $ is the percent of familiarity of the root, $ \frac{1}{m}\sum_{j=1}^mP_f(k_{j},t) $ calculates the average percent of familiarities of its descendants (not including the root). E.g., Figure 9 is the Understanding Tree of Figure 8 with Familiarity Measures transformed into percentages (assuming $ f_{T} $ equals 100). The $ P_f(k_{r},t) $ of it equals 85\%, and $ \frac{1}{m}\sum_{j=1}^mP_f(k_{j},t) $ equals 89\%, so the $ P_u(k_{r},t) $ equals 76\%, indicating the subject's understanding of the root Knowledge Point is 76\%.  

If $ P_u(k_{r},t) $ is less than 100\%, the Knowledge Point is classified as ``Not Understood". Thus it can be seen that we are using a conservative strategy for estimating the subject's understanding. For a Knowledge Point to be classified as ``Understood", the subject must be familiar with \emph{every} node of its Understanding Tree.

\begin{equation}  
	P_f(k_{i},t)=\left\{
	\begin{array}{cl}
		 1                  &    f(k_{i},t) \geq f_{T} \\
		 f(k_{i},t)/f_{T}   &    f(k_{i},t) < f_{T}
	\end{array}
	\right.  
\end{equation}

\begin{equation}
	P_u(k_{r},t) = P_f(k_{r},t)*\frac{1}{m}\sum_{j=1}^m P_f(k_{j},t)
\end{equation}

If $ P_u(k_{r},t) $ equals 100\%, the subject is assumed having understood the root Knowledge Point, then the average Familiarity Measure of the Understanding Tree divided by the threshold features the magnitude of understanding. Since people usually are very familiar with the BKPs, it may be preferable to exclude them or normalize their effects when computing the average Familiarity Measure. 

Another choice for calculating $ P_f(k_{r},t) $ or $ P_f(k_{i},t)$ is to use a sigmoid function.
\begin{equation*}  
P_f(k_{i},t)={\frac {1}{1+e^{-a(f(k_{i},t)- b)}}}
\end{equation*}
Where $ f(k_{i},t)$ is the subject's Familiarity Measure of Knowledge Point $k_{i}$ at time $t$, $a$ and $b$ are two constant parameters to be optimized.

\subsection{Construction of Understanding Tree}
An Understanding Tree can be constructed manually by a group of experts, or generated automatically by machines. Algorithm 2 shows the steps to generate an Understanding Tree automatically. Table 3 illustrates three definitions of the Central Limit Theorem (CLT)\footnote{They can be found at \url{http://www.math.uah.edu/stat/sample/CLT.html}, \url{http://sphweb.bumc.bu.edu/otlt/MPH-Modules/BS/BS704_Probability/BS704_Probability12.html}, \url{http://stattrek.com/statistics/dictionary.aspx}}, the third column lists the involved Knowledge Points in each definition. According to the rule mentioned in Algorithm 2, the child nodes of CLT are selected as \emph{sample}, \emph{distribution}, \emph{mean}, \emph{independent}, and \emph{normal} (Knowledge Points that are not BKPs can be further extended). To be safe, the generated Understanding Tree can be inspected by human experts. Understanding Tree is a static structure, once constructed, it is not likely to change them.

\begin{algorithm} 
	\caption{ An algorithm to construct Understanding Tree}  
	\begin{algorithmic}[1] 
		\REQUIRE ~~\\ 
		
		The set of all Knowledge Points, $ \Omega $;\\
		The set of all BKPs, $ B $;\\
		The root Knowledge Point, $ R_k $;
		
		\ENSURE ~~\\ 
		$ R_k $'s Understanding Tree;
		\STATE Search definitions of $ R_k $ from a library of authoritative documents;
		\STATE Discover involved Knowledge Points for each definition;
		\STATE Select Knowledge Points according to some rules, e.g., more than half of the definitions have referenced the Knowledge Point; 
		\STATE Recursively extend non-BKP Knowledge Points that are selected;
		\RETURN All the selected Knowledge Points;		
	\end{algorithmic}
\end{algorithm} 

\begin{table*}
	\centering
	\begin{tabular}{|C{1cm}|L{12cm}|C{3.5cm}|}
		\hline
		& \qquad \qquad \qquad \qquad \qquad \qquad \qquad \qquad Content & Knowledge Points  \\ \hline
		\qquad \qquad \qquad \qquad \qquad \qquad 1 & The Central Limit Theorem (CLT) states that the sampling distribution of the mean of any independent, random variable will be normal or nearly normal, if the sample size is large enough. &  CLT,  sample, distribution, mean, independent, random variable, normal, size \\ \hline
		\qquad \qquad \qquad \qquad \qquad \qquad 2 & The Central Limit Theorem (CLT) states that the distribution of the sum (or average) of a large number of independent, identically distributed variables will be approximately normal, regardless of the underlying distribution.  &  CLT,  distribution, sum,  average, independent, variable, normal \\ \hline
		\qquad \qquad \qquad \qquad \qquad \qquad 3 &  The Central Limit Theorem (CLT) states that if you have a population with mean $ \mu $ and standard deviation $ \sigma $ and take sufficiently large random samples from the population with replacement, then the distribution of the sample means will be approximately normally distributed.  &  CLT,  population, standard deviation, random, replacement, distribution, sample, mean, normal \\  
		\hline 
	\end{tabular}
	\caption{Three definitions of the Central Limit Theorem (CLT)}
\end{table*}

\section{Facilitate meaningful learning}
Knowing one's understanding degrees of knowledge helps practice meaningful learning. Meaningful learning is the concept that learned knowledge is fully understood to the extent that it relates to other knowledge, implies there is a comprehensive knowledge of the context of the facts learned \cite{Mayer2002}. ICKEM helps people practice Computer-aided Incremental Meaningful Learning (CAIML). CAIML is defined as a strategy of letting the computer estimate a person's current knowledge states, then recommend the optimum learning material for the subject to accomplish a meaningful learning. The optimum material introduces some new knowledge blended with old knowledge the subject has known, meanwhile, interpreting the new knowledge.

For example, a college student, who has comprehended the basic knowledge of Advanced Mathematics and Computer Science, wants to be an expert of Artificial Intelligence (AI) by teaching himself. A professor recommends him 1,000 documents related to AI, and asserts if he can fully understand the contents of the documents, he will be an expert of AI. The question is: what is the best sequence for the student to learn the documents? Some documents are easy to understand, and should be read in the beginning; some documents are intricate, learning them in the first place is painful and frustrating, and should be put at the end of the learning process. If a computer knows a person's knowledge states at any time, it is not difficult to make the learning plan, and recommend the optimum document for current learning.
\subsection{An algorithm to facilitate CAIML}
Algorithm 3 is devised to facilitate CAIML. It recommends the optimum document for the subject to learn, by analyzing his current knowledge states. It searches for the document which has the least Knowledge Points that are not understood by the subject, which implies it is the easiest document to understand at present. In the document, the ``Understood" Knowledge Points server as interpretations to the ``Not Understood" ones when learning the document. Therefore, the algorithm facilitates a meaningful learning.

\begin{algorithm}
	\caption{ An algorithm to recommend the optimum document(s) for CAIML}  
	\begin{algorithmic}[1] 
		\REQUIRE ~~\\ 
		
		The set of documents to be learned;\\
		The set of all Knowledge Points, tagged with the subject's Familiarity Measures to them;\\
		The set of all non-BKP Knowledge Points' Understanding Trees;
		
		\ENSURE ~~\\ 
		The optimum document(s) for current learning to practice a meaningful learning;
		\STATE Extract involved Knowledge Points in each document, classify them as ``Understood" and ``Not Understood", according to the rule mentioned in Section 3.1;
		\STATE Count the number of ``Not Understood" Knowledge Points for each document;
		\RETURN The document(s) with the least ``Not Understood" Knowledge Points;		
	\end{algorithmic}
\end{algorithm}

An alternative method to recommend the optimum document is to estimate the subject's anticipated percentage of understanding of each document, then return the one that is most approaching 100\%. A person's understanding of a document is calculated with Equation 9. Suppose the document involves $ n $ Knowledge Points, $ P_u(d,t) $ is the percent of understanding of the document at time $ t $; $ \xi_{i} $ is Knowledge Point $ k_{i} $'s share of the document, its calculation has been discussed in Section 2.4; $ P_u(k_{i},t) $  is Knowledge Point $ k_{i} $'s percent of understanding at time $ t $. If a person has understood all the Knowledge Points the document contains, its $ P_u(d,t) $ is 100\%. It is possible that a person learns a ``100\% understood" document to strengthen his knowledge. As Equation 9 can estimate a person's understanding degree to document, it can also be used to recommend appropriate reviewers for a paper.

\begin{equation}
	P_u(d,t)  =  \sum_{i=1}^n\xi_{i}*P_u(k_{i},t)
\end{equation}

People's memory fades away over time, the Familiarity Measures decrease accordingly. It is also possible to recommend a document that had been tagged ``100\% understood", because the $ P_u(d,t) $ has abated.
\subsection{An example of CAIML}
Here is an example to illustrate the logic of CAIML. Suppose a person wants to fully understand the suite of documents listed in Table 2 by learning them. Figure 7 shows all the Knowledge Points involved in the documents, the third column of Table 2 lists ones referenced by each of them. The subject is assumed to have understood the BKPs before the beginning of the learning, which are the leaf nodes of Figure 7 (if not, he can learn them first). Table 4 shows the number of ``Not Understood" Knowledge Points before the beginning of the learning and after each learning, for each document. E.g., there are 8 ``Not Understood" Knowledge Points before the starting of the learning for D1, they are SSP, SP, JPD, PM, SS, RV, PS, and PD. According to Algorithm 3, the subject should first learn either of D5, D7 or D8. After learning of D5, the subject is assumed to have understood RV. Therefore, the number of ``Not Understood" Knowledge Points for D1 becomes 7. The first column of Table 4 suggests one of the optimum learning sequences of the documents, which is D5, D8, D4, D2, D7, D6, D3, and D1.

\begin{table}
	\centering
	\begin{tabular}{ c  c  c  c  c  c  c  c  c }
		\hline
		& D1 & D2 & D3 & D4 & D5 & D6 & D7 & D8\\ \hline
		Before starting & 8 & 3 & 5 & 2 & 1 & 2 & 1 & 1\\ \hline
		D5 & 7 & 3 & 4 & 2 & 0 & 2 & 1 & 1\\ \hline
		D8 & 6 & 2 & 3 & 1 & 0 & 1 & 1 & 0\\ \hline
		D4 & 5 & 1 & 3 & 0 & 0 & 1 & 1 & 0\\ \hline
		D2 & 4 & 0 & 3 & 0 & 0 & 1 & 1 & 0\\ \hline
		D7 & 3 & 0 & 2 & 0 & 0 & 1 & 0 & 0\\ \hline
		D6 & 2 & 0 & 1 & 0 & 0 & 0 & 0 & 0\\ \hline
		D3 & 1 & 0 & 0 & 0 & 0 & 0 & 0 & 0\\ \hline
		D1 & 0 & 0 & 0 & 0 & 0 & 0 & 0 & 0\\   	 
		\hline 
	\end{tabular}
	\caption{An example of CAIML}
\end{table}

\section{Discussion}
Assessing a person's understanding of conceptual knowledge is not easy; as an experimental model aimed to accomplish this, a great deal of thought and research are required to realize its potential.

\subsection{Dealing with similar Knowledge Points}
There is a problem of how to deal with Knowledge Points that are different but have little distinction, such as ``random variation" and ``random variable". One solution is to homogenize them to the same Knowledge Point; another is to compensate one Knowledge Point's Familiarity Measure with others' Familiarity Measures, because learning others helps to understand it. The compensation can be calculated with Equation 10. $ F_{k_{i}} $ is Knowledge Point $ k_{i} $'s Familiarity Measure, each of its sibling contributes $ 1/c_{j} $ of its Familiarity Measure to $ k_{i} $ ($ c_{j} $ is coefficient to be optimized). 

\begin{equation}
	F_{k_{i}\_new} = F_{k_{i}\_old} + \sum_{j=1}^m{\frac{1}{c_{j}}}F_{k_{j}}
\end{equation}

\subsection{Trade-offs of evaluating one's understanding of conceptual knowledge}
Quantitatively assessing one's understanding of conceptual knowledge seems to be a good thing, but there are risks that it introduces some harmful effects. For example, if the Familiarity Measures and understanding degrees calculated are inaccurate, it may lead to wrong decisions. In addition, it cannot detect a person's talent and potential in a field. On the other hand, traditional exams or interviews have their limitations. For example, it needs other people's cooperation to accomplish the evaluation; it only assesses one's knowledge in a particular field at a time, and the evaluation is not comprehensive. Since it only assesses questions being asked, not all of the topics in a field are evaluated. ICKEM assesses one's knowledge independently, comprehensively, and automatically. Therefore, the methods of evaluating one's understanding of knowledge should be used cooperatively, complementing one another.

\subsection{Privacy issues}
Recording one's learning history will inevitably violates privacy. To protect privacy, the learning histories can be password protected or encrypted and stored in personal storage; they should not be revealed to other people. The only information that can be viewed by the outside world is the individual's knowledge measures of some Knowledge Points. The Knowledge Points that may involve privacy are separated from others; every output of them should be authorized by the owner. A person may choose to keep all of his knowledge measures private.

\subsection{Analyzing  with topic models}
Since we can calculate a person's Familiarity Measures to different Knowledge Points. The Familiarity Measures can be considered as term frequencies in a document. Therefore, at a given time,  a person is equivalent to a document. Thus we can use probabilistic topic models \cite{hofmann1999probabilistic,blei2003latent} to analyze multiple people's expertise of knowledge, or using Understanding Map Supervised Topic Model (UM-S-TM) \cite{liu2021topic} to analyze an individual.

\subsection{Estimating with supervised learning models}
If we have a training data set, which tags whether a person has understood a Knowledge Point at a given time $t$, through traditional methods like tests or interviews, then we can train a supervised machine learning model to predict if a person has understood a Knowledge Point at time $t$, such as using logistic regression, or neural networks. The input information is the subject's average Familiarity Measures to different levels of Knowledge Points on an Understanding Tree.

\section{Related work}
Many research fields focus on the collection of personal information, such as lifelogging, expertise finding, and personal informatics. Bush envisioned the `memex' system, in which individuals could compress and store personally experienced information, such as books, records, and communications \cite{bush1979we}. Inspired by `memex', Gemmell et al. developed a project called MyLifeBits to store all of a person's digital media, including documents, images, audio, and video \cite{gemmell2002mylifebits}. In \cite{Liu2017}, a person's reading history about an electronic document is used as attributes for re-finding the document. ICKEM is similar to `memex' and MyLifeBits in that it records an individual's digital history, although for a different purpose. `Memex' and MyLifeBits are mainly for re-finding or reviewing personal data; ICKEM is for quantitatively evaluating a person's knowledge.

Personal informatics is a class of tools that help people collect personally relevant information for the purpose of self-reflection and gaining self-knowledge \cite{li2010stage,wolf2009know,yau2009self}. Various tools have been developed to help people collect and analyze different kinds of personal information, such as location \cite{lindqvist2011m}, finances \cite{kaye2014money}, food \cite{cordeiro2015barriers}, weight \cite{kay2013there,maitland2011designing}, and physical activity \cite{fritz2014persuasive}. ICKEM facilitates a new type of personal informatics tool that helps people discover their expertise and deficiencies in a more accurate way, by quantitatively assessing an individual's understanding of knowledge.

Expertise is one's expert skill or knowledge in a particular field. Expertise finding is the use of tools for finding and assessing individual expertise \cite{mcdonald1998just,mattox1999enterprise,vivacqua1999agents}. As an important link of knowledge sharing, expertise finding has been heavily studied in many research communities \cite{ackerman2013sharing, Cheng2014,maybury2002awareness,tang2008arnetminer,Aslay2013,guy2013mining}. Many sources of data have been exploited to assess an individual's expertise, such as one's publications, documents, emails, web search behavior, other people's recommendations, social media etc. ICKEM provides a new source of data to analyze one's expertise--one's learning history about a topic, which is more comprehensive and straightforward than other data sources. Because one's expertise is mainly obtained through learning (Including ``Informal Learning", which occurs through the experience of day-to-day situations, such as a casual conversation, play, exploring, etc.)

\section{Conclusion}
People's pursuing of knowledge is never stopping. Most conceptual knowledge is transmitted through language; it is hard to imagine how a person can obtain conceptual knowledge without using language. A piece of written or spoken language can be converted into text.  We propose a new method to estimate a person's understanding of a piece of conceptual knowledge, by analyzing the text content of one's all learning experiences about a knowledge topic. The computation of familiarity degree takes into account the total time the subject has devoted to a knowledge topic, a topic's share in a learning session, the subject's physical and psychological status during a session, the memory decay of each learning experience over time, and the difference among learning methods. To estimate a person's understanding degree to a knowledge topic, it comprehensively evaluates one's familiarity degrees to the topic itself and other topics that are essential to understand the topic.
Quantitatively evaluating a person's understanding of knowledge facilitates many applications, such as personalized recommendation, meaningful learning, expertise and deficiency finding etc. With the prevailing of wearable computers like Google Glass and Apple Watch, and maturing of technologies like Speech Recognition and Optical Character Recognition (OCR), it is practicable to analysis people's daily learning activities like talking, listening, and reading. Therefore, ICKEM is technically feasible.

\bibliographystyle{abbrv}
\bibliography{kmodel}

\balance

\end{document}